\documentclass[%
reprint,
 amsmath,amssymb,
 aps,
 superscriptaddress,
]{revtex4-2}

\usepackage{graphicx}% Include figure files
\usepackage{dcolumn}% Align table columns on decimal point
\usepackage{bm}% bold math
\usepackage{hyperref}% add hypertext capabilities
\usepackage{xcolor}
\hypersetup{
    colorlinks,
    linkcolor     = {red!30!black},
    citecolor     = {blue!30!black},
    urlcolor      = {green!20!black},
}
\usepackage{physics}
\usepackage{bbold}
\usepackage{amsmath}
\usepackage{amsfonts}
\usepackage{amssymb}
\usepackage{mathtools}
\usepackage{tensor}

\newcommand{\corr}[1]{#1}  

\usepackage{placeins}

\begin{document}
\title{A Variational Ansatz for the Ground State of the Quantum Sherrington-Kirkpatrick Model}
\author{Paul M. Schindler}
    \email{psch@pks.mpg.de}
    \affiliation{Max-Planck-Institut f\"ur Physik komplexer Systeme, N\"othnitzer Str.~38, 01187 Dresden, Germany}
    \affiliation{Max-Planck-Institut für Quantenoptik, Hans-Kopfermann-Str.~1, 85748 Garching, Germany}
\author{Tommaso Guaita}
    \email{tommaso.guaita@fu-berlin.de}
    \affiliation{Max-Planck-Institut für Quantenoptik, Hans-Kopfermann-Str.~1, 85748 Garching, Germany}
    \affiliation{Munich Center for Quantum Science and Technology, Schellingstr.~4, 80799 München, Germany}
    \affiliation{Dahlem Center for Complex Quantum Systems, Freie Universit\"at Berlin, Arnimallee 14, 14195 Berlin, Germany}
\author{Tao Shi}
\affiliation{CAS Key Laboratory of Theoretical Physics, Institute of Theoretical Physics, Chinese Academy of Sciences, Beijing 100190, China}
\affiliation{CAS Center for Excellence in Topological Quantum Computation, University of Chinese Academy of Sciences, Beijing 100049, China}

\author{Eugene Demler}
\affiliation{Institute for Theoretical Physics, ETH Zurich, Wolfgang-Pauli-Str. 27, 8093 Zurich, Switzerland}

\author{J. Ignacio Cirac}
    \affiliation{Max-Planck-Institut für Quantenoptik, Hans-Kopfermann-Str.~1, 85748 Garching, Germany}
    \affiliation{Munich Center for Quantum Science and Technology, Schellingstr.~4, 80799 München, Germany}

% Abstract

\begin{abstract}
We present an ansatz for the ground states of the Quantum Sherrington-Kirkpatrick model, a paradigmatic model for quantum spin glasses. Our ansatz, based on the concept of generalized coherent states, very well captures the fundamental aspects of the model, including the ground state energy and the position of the spin glass phase transition. It further enables us to study some previously unexplored features, such as the non-vanishing longitudinal field regime and the entanglement structure of the ground states. We find that the ground state entanglement can be captured by a simple ensemble of weighted graph states with normally distributed phase gates, leading to a volume law entanglement, contrasting with predictions based on entanglement monogamy.

\end{abstract}
\maketitle

% begin Introduction

%1
\textit{Introduction} --
Spin glasses are an important paradigm in statistical physics. Besides their relevance in describing disordered classical magnets~\cite{SpinGlass_overview,nishimori01}, it was shown that optimization tasks, such as the traveling salesman problem, can be mapped to solving for the ground states of spin glass systems~\cite{Jesi_2016_history_SG,information_book,SpinGlass_overview}.
Classical spin glasses can be promoted to quantum models by introducing a transverse field. The resulting quantum spin glasses form by themselves an important playground to study the interplay of disorder and frustration with quantum effects~\cite{Sachdev_QSG}. Moreover, there is evidence that the quantumness can be exploited to shortcut optimization tasks, for instance through quantum annealing~\cite{santoro_theory_2002,adiabatic_qc,quantum_optimization,quantum_opt_qspinglass,Farhi2019TheQA}.

%2
The textbook example of a quantum spin glass model is the Quantum Sherrington-Kirkpatrick~(QSK) model, a generalization of the classical Sherrington-Kirkpatrick~(SK) model~\cite{sherrington_solvable_1975, Ishii_1985_first_time_QSK}. The QSK model has been studied extensively in the literature both analytically~\cite{Ishii_1985_first_time_QSK,Usadel_1986,Thirumalai_1989,Yamamoto_SK1.5,Takahashi_FieldTheory,Goldschmidt_Lai_1990,miller_zero-temperature_1993} and numerically~\cite{Alvarez_1996,Lai_1990,Chakrabarti_noRSB,chakrabarti_1996,Das_2008,Mukherjee_ED_MC,Takahashi_EnergyGap,Koh_EnergyGap,Mukherjee_2018_MBL_SK,Mukherjee_Ergodicity,young_stability_2017,Pappalardi_2020_semiclassical}. While the famous Parisi solution~\cite{Parisi_RSBsolution,SK_overview} provides a full solution to the classical SK model, many open questions remain for the quantum SK model.
Since the QSK model is an all-to-all coupled model one might assume that a mean-field product state ansatz well describes the ground state. However, this ansatz predicts a quantum phase transition~(QPT) from a quantum spin glass phase to a paramagnetic phase at a critical transverse field $g_\mathrm{C}\approx 2\,J$~\cite{Koh_EnergyGap}. 
Field theory approaches~\cite{Yamamoto_SK1.5,Takahashi_FieldTheory,miller_zero-temperature_1993} using the replica method suggest instead a phase transition at $g_\mathrm{C}\approx 1.5\,J$. 
Numerical calculations at small system sizes~\cite{arrachea_dynamical_2001,Mukherjee_ED_MC,Das_2008} or obtained at finite temperature~\cite{Alvarez_1996,Lai_1990,rozenberg_dynamics_1998,Mukherjee_ED_MC,young_stability_2017} confirm the latter~\cite{factor2}. 
So far no good ansätze have been found which can describe the zero temperature regime for large system sizes, preventing the study of further properties of the ground state, such as entanglement. 

%3
Here, we consider a variational family, motivated by the concept of generalized group-theoretic coherent states~\cite{Guaita_etal_2021}, which extends the product state ansatz introducing a richer entanglement structure. The special structure of these states allows us to introduce non-trivial quantum correlations while preserving the ability to efficiently compute variational ground states up to large system sizes of $N=200$ spins. 
We additionally develop a method to study the entanglement structure of the ground states.
Our results show a volume law of entanglement, which indicates that entanglement monogamy does not provide a scaling constraint despite the fact that the QSK model involves all-to-all spin interactions. Furthermore, this entanglement structure is also identified within a set of states that have been introduced in the Quantum Information context, namely weighted graph states~\cite{hartmann_weighted_2007} with normally distributed random phase gates.

%4
\textit{The model} --
Concretely, the QSK model corresponds to a mixed field Ising model with all-to-all couplings between the $N$ spins and quenched disorder in the couplings and longitudinal field,
\begin{equation}
    H_\mathrm{QSK} = - \frac{1}{2} \sum_{n,\, m=1}^N J_{nm} \sigma_n^z \sigma_m^z - g \sum_{n=1}^N \sigma_n^x - \sum_{n=1}^N h_n \sigma_n^z \, , \label{eq:QSK_model}
\end{equation}
where $\sigma_n^k$ is the $k$'th Pauli-matrix acting on the $n$'th spin. The longitudinal field $h_n$ and the couplings $J_{nm}$ are independently normally distributed numbers with zero mean and variance $\overline{h_n^2}=h^2$ and $\overline{J_{nm}^2} = J/N $, respectively.
Here and in the following we use the convention that an overbar $\overline{\:\vphantom{i}\cdot\:}$ indicates disorder average and we will mostly concentrate on the case $h=0$.

%5
\textit{Variational ansatz} -- 
Our variational ansatz was first introduced in Refs.~\cite{anders_ground-state_2006, anders_variational_2007}. It generalizes the ansatz of atomic coherent states~(CS)~\cite{arecchi_atomic_1972},
\begin{equation}
    \ket{\phi\pqty{\boldsymbol{x}}} = \mathcal{U}\pqty{\boldsymbol{x}} \ket{\uparrow,\, \dots,\, \uparrow} \, ,\label{eq:cs}
\end{equation}
where $\sigma^z \ket{\uparrow}=+\ket{\uparrow}$ and $\mathcal{U}\pqty{\boldsymbol{x}} = \exp(-i \sum_{n,k} x_n^k \sigma_n^k )$ rotates each of the $N$ spins individually on the Bloch sphere. The CS ansatz is parametrized by $x_n^k\in\mathbb{R}$ and corresponds to the set of normalized product states.

%6
A generalization procedure~\cite{Guaita_etal_2021} leads to generalized atomic coherent states~(GCS)
\begin{equation}
    \ket{\Psi\pqty{\boldsymbol{x},\, \boldsymbol{y},\, \boldsymbol{M}}} = \mathcal{U}\pqty{\boldsymbol{y}} \mathcal{V}\pqty{\boldsymbol{M}} \ket{\phi\pqty{\boldsymbol{x}}}\, , \label{eq:gcs}
\end{equation}
where $x_n^k$, $y_n^k$ and $M_{nm}$~($n<m$) are the variational parameters. $\mathcal{U}$ and $\ket{\phi}$ are defined as in equation~\eqref{eq:cs} and the entangling unitary $\mathcal{V}\pqty{\boldsymbol{M}}$ is given by
\begin{equation}
    \mathcal{V}\pqty{\boldsymbol{M}}=\exp(-\frac{i}{4}\sum_{n<m} M_{nm}\sigma_n^z \sigma_m^z)\,,
\end{equation}
for any real symmetric matrix $\boldsymbol{M}$.

%7
The entangling unitaries $\mathcal{V}\pqty{\boldsymbol{M}}$ contain two-spin terms which give the states~\eqref{eq:gcs} a non-trivial correlation structure. Nonetheless, when computing expectation values of Pauli operators we have $\mathcal{V}\pqty{\boldsymbol{M}}^\dagger \sigma_n^\pm \mathcal{V}\pqty{\boldsymbol{M}} = \sigma_n^\pm \exp(\pm i/4 \sum_m M_{nm} \sigma_m^z)$, that is the two-spin terms cancel and we are left just with products of single spin operators~\cite{Foss_Feig_2013_Noneq,Guaita_etal_2021}. This crucial property allows us to find analytical expressions for the energy and the gradient of the energy with respect to the variational parameters~\cite{appendixA}. Thanks to this, we can efficiently obtain the variational ground states of individual Hamiltonian realizations for large system sizes of up to $N=200$ spins through a natural gradient descent algorithm~\cite{hackl2020geometry}.

\begin{figure}
    \centering
    \includegraphics[width=0.5\textwidth]{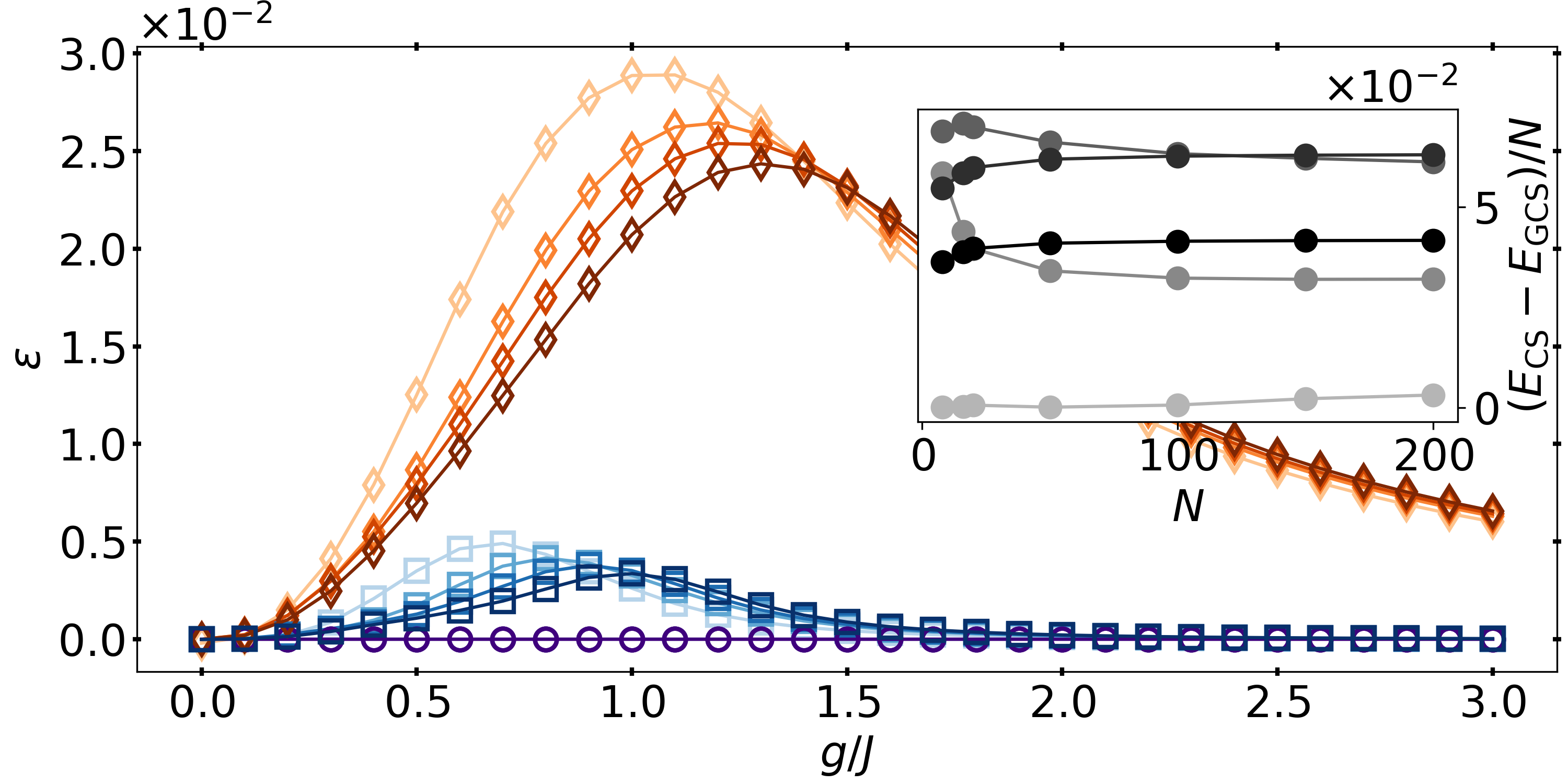}
    \caption{Average error in energy density $\varepsilon=\overline{\Delta E}/\overline{W}$ as a function of the transverse field for different methods~(CS in orange, GCS in blue and ED in purple) and system sizes $N=8,\,12,\,16,\,22$~(light to dark). $\Delta E$ is the difference between the variational energy and the exact ground state energy, W is the difference between the highest and lowest energies in the exact spectrum.
    \textit{Inset:} Difference between CS and GCS energies per site $(E_\mathrm{CS} - E_\mathrm{GCS})/N$ as a function of the system size $N$ for different values $g/J=0.1,\, 1.0,\, 1.5,\, 2.0,\, 3.0$~(light to dark).
    All data is for $h=0$ and averaged over $n_\mathrm{samples}=1000$ disorder realizations.}
    \label{fig:energy}
\end{figure}

%8
To demonstrate the expressivity of the GCS ansatz, we first consider the approximate ground state energy.
For small system sizes we can compare the variational energies with numerically exact results, obtained via a Lanczos Exact Diagonalization method~(ED)~\cite{Lanczos_1950}, see Fig.~\ref{fig:energy}. 
We find good quantitative agreement of the variational energy with the exact ground state energy over a broad range of transverse and longitudinal field values. In particular a notable improvement of the GCS ansatz upon the CS ansatz becomes visible. The method performs worst in a region with $0.5\,J<g<1.5\,J$. For the assessed system sizes, the point of maximal error moves with growing $N$ towards the expected critical point $g_\mathrm{C} \approx 1.5\,J$, while the maximal error value decreases~\cite{bump}. This agreement is not limited to the energy but can also be seen for other observables of interest~\cite{observables}.

For larger systems it is no longer possible to compare to an exact solution. However we observe an extensive improvement in energy upon the CS ansatz, suggesting that the GCS ansatz gives a non-vanishing improvement even in the thermodynamic limit, see inset of Fig.~\ref{fig:energy}.

\begin{figure}
    \centering
    \includegraphics[width=0.5\textwidth]{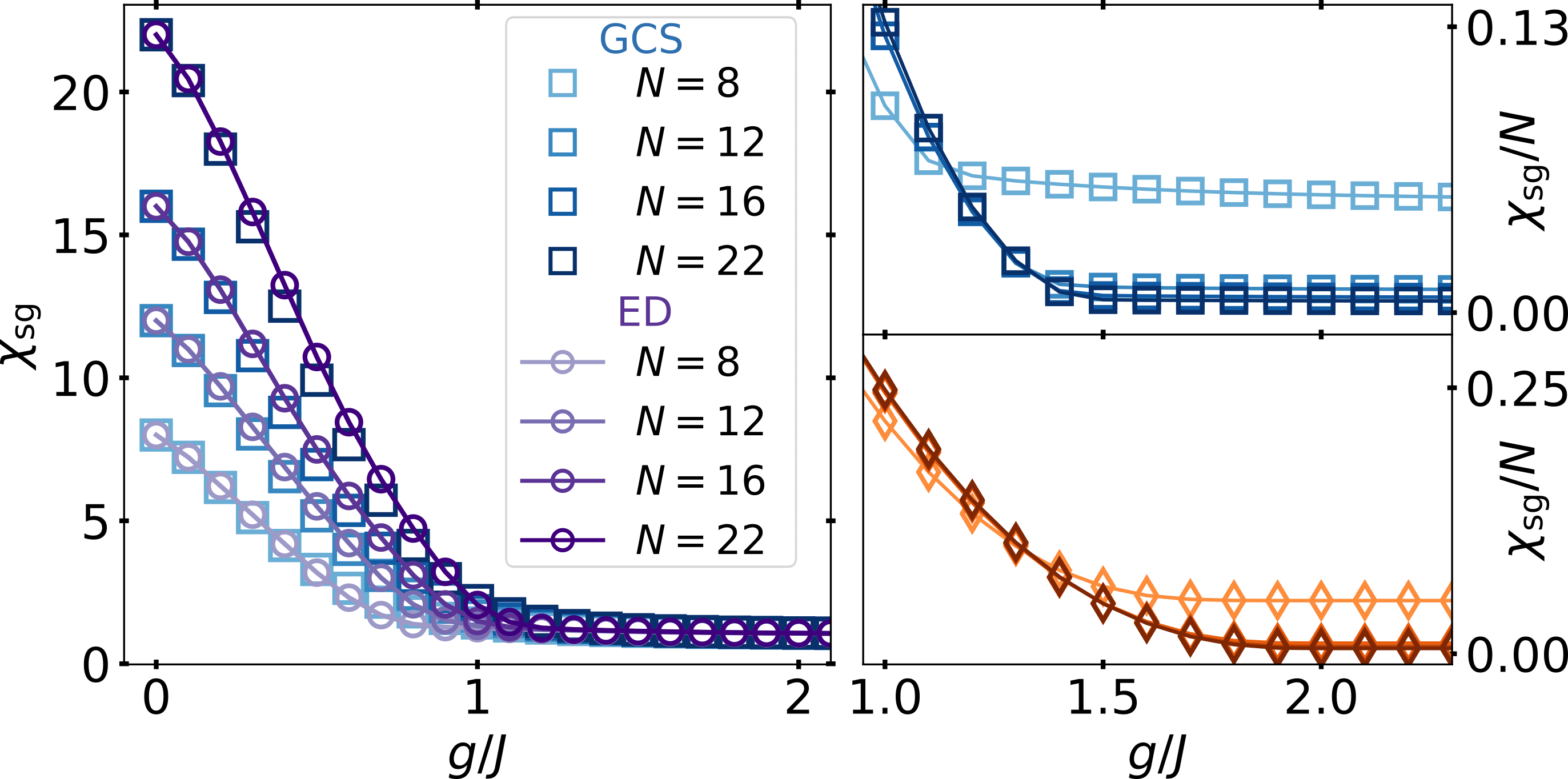}
    \caption{\textit{Left Panel:} Spin glass susceptibility $\chi_\mathrm{sg}$ as a function of the transverse field $g$ for ED~(purple circles) and GCS~(blue squares).
    \textit{Right Panels:} Spin glass susceptibility per site $\chi_\mathrm{sg}/N$ for GCS~(top, blue) and CS~(bottom, orange) and system sizes $N=20,\, 100,\, 200$~(from light to dark).
    All data is for $h=0$ and averaged over $n_\mathrm{samples}=1000$ disorder realizations.}
    \label{fig:order_parameter}
\end{figure}

%8b
\textit{Quantum phase transition} --
Our variational ansatz also allows us to study the QPT on the $h=0$ line of the model's parameter space. 
For this we consider the spin glass susceptibility $\chi_\mathrm{sg} = N^{-1} \sum_{n,m} \overline{\expval{\sigma_n^z \sigma_m^z}^2}$. Indeed, the susceptibility per site $\chi_\mathrm{sg}/N$, which is independent of the system size in the thermodynamic limit, vanishes in the paramagnetic phase (large $g$) and is finite in the spin glass phase (small $g$)~\cite{SpinGlass_overview,sen1997quantum,Mukherjee_2018_MBL_SK,Das_2008,Leschke_2021}.
For small system sizes we find good quantitative agreement of the variational value of the susceptibility with numerically exact~(ED) results, see left panel of Fig.~\ref{fig:order_parameter}. 
More importantly, the variational ansatz enables us to study the system at much larger sizes, see right panels of Fig.~\ref{fig:order_parameter}. For $N\geq 100$ finite size effects are almost absent, allowing us to read off the critical field value directly.
Both variational ansätze clearly indicate the existence of a phase transition. However, in agreement with the literature~\cite{Koh_EnergyGap}, the CS underestimate the quantum fluctuations showing a phase transition at roughly $g_\mathrm{C}\approx 2\,J$. In contrast, the GCS capture the true critical point at $g_\mathrm{C}\approx 1.5\,J$.  Thus, the additional entanglement structure introduced in the GCS not only leads to an improvement in energy but also seems crucial in capturing the physics of the QSK model in the thermodynamic limit. This agreement with established results further indicates that the variational states remain a good approximation of the ground states in the large $N$ limit.

%9
\textit{Entanglement structure of the ground state} --
The findings above suggest that the GCS ansatz describes the ground state of the QSK model very well for all system sizes up to the thermodynamic limit.
Having such an explicit expression for the ground state wavefunction allows us to study in detail its entanglement properties. Before looking into the numerical results, we will consider some hypotheses about the expected entanglement behaviour.

%10
First, let us try to understand the role of the additional two-spin entangling gates contained in $\mathcal{V}\pqty{\boldsymbol{M}}$ by taking a closer look at the matrix elements $M_{nm}$. %, see fig.~\ref{fig:Mnm}. 
Considering the probability distribution $p\pqty{M_{nm}}$ over many disorder realizations, we observe that it resembles a Gaussian distribution with zero mean and variance scaling as $1/N$. 
In addition, we find that the mean level spacing ratio averaged over many realizations yields $\overline{\expval{r}}\approx0.53$ roughly independent of the transverse field value $g>0$ and system size $N$, which is in agreement with the result of the Gaussian Orthogonal Ensemble~(GOE)~\cite{Pal_levelratio_2010}.

%11
This implies that most two-spin entangling gates approach the identity as $N\to \infty$.
This may seem compatible with the naive hypothesis that, due to the mean-field nature of the model, product states should well describe the ground state, at least in the thermodynamic limit. This assumption would predict the entanglement entropy between any two subsystems going to zero as $N\to\infty$.

%12
Note, however, that the number of entangling gates acting on each individual spin diverges in this limit, suggesting that a non-trivial entanglement structure is still possible.
Indeed, let us consider a subsystem $A$ composed of the first $L$ spins. We quantify the entanglement between these $L$ spins and the rest of the system by computing the second R\'enyi entropy $S_2(L)$ of the subsystem's reduced density matrix. Given the all-to-all connectivity of our ansatz, there exist $L(N-L)$ two-spin entangling gates acting between spins in $A$ and in its complement $A^c$. Each of these gates individually generates a two-spin state with an average entanglement entropy proportional to $\overline{M_{nm}^2}\sim 1/N$.  The cancellation of these two scalings could suggest a second hypothesis, i.e. that the entanglement entropy between $A$ and $A^c$ is proportional to $L$ in the thermodynamic limit $N\to\infty$. This expectation can also be made more rigorous with an argument based on the Central Limit Theorem~\cite{appendixA}.

\begin{figure}
    \centering
    \includegraphics[width=0.5\textwidth]{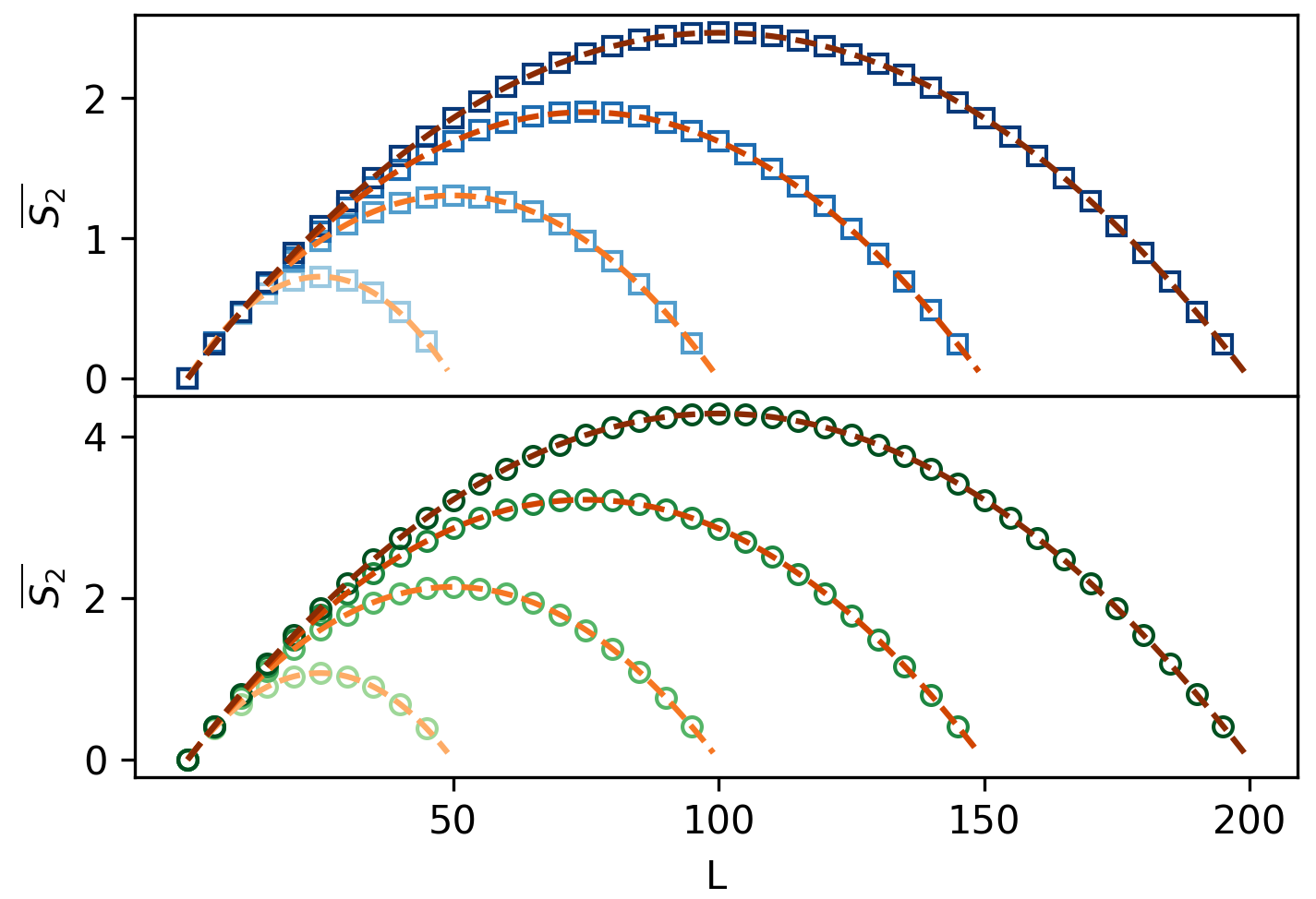}
    \caption{Average Renyi-2 entanglement entropy as a function of the subsystem size $L$ for the QSK ground state at $g=1\, J$, $h=0$ (top panel in blue) and for an ensemble of weighted graph states~\eqref{eq:wgs} (bottom panel in green). Data is plotted for total system sizes $N=50,100,150,200$ (from light to dark markers). In all cases, including ground state data for other values of the fields $g$ and $h$, the entropy is well fitted by the function~\eqref{eq:entanglement-fit} (orange dashed lines).
    }
    \label{fig:entropy_fit}
\end{figure}

%13
As a third alternative, we may compare the model to a related but analytically solvable model, namely a model with all-to-all interactions and invariant under spin permutations. Notice that in our case, due to the disordered nature of the QSK model, individual realizations of the couplings $J_{nm}$ and $h_n$ are not permutationally invariant. However, invariance is present upon disorder averaging, so the permutationally invariant case may still provide a useful comparison. In such case the ground state $\ket{\Psi}$ must possess a Schmidt decomposition
\begin{equation}
    \ket{\Psi}=\sum_k \lambda_k \ket{\varphi_k}\ket{\eta_k}\,,
    \label{eq:schmidt-decomp}
\end{equation}
where $\ket{\varphi_k}$ and $\ket{\eta_k}$ are orthonormal states of $A$ and $A^c$ respectively. Due to the symmetry, the states $\ket{\varphi_k}$ must in particular belong to the subspace of permutationally invariant states of $A$. Such subspace has dimension $L+1$, so there can be at most $L+1$ terms in the sum~\eqref{eq:schmidt-decomp}. It follows that the entanglement entropy of $A$ is bounded by $S_2(L)\leq \log(1+L)$.
This scaling of the entanglement can  be viewed as a consequence of entanglement monogamy~\cite{koashi_monogamy_2004,terhal_is_2004}.

%14
We would like now to compare our results with these hypotheses. 
To this end, we have developed an efficient method to numerically compute $S_2(L)$ for the states~\eqref{eq:gcs}, reducing the problem to the one of
estimating averages for a classical sampling problem~\cite{appendixA}.
%computing averaged properties of a related classical model using Monte Carlo methods~\cite{appendixA}.
%To do this we have computed numerically the average entanglement entropy $\overline{S_2(L)}$ for our variational ground states as a function of $L$ for varying system sizes $N$~\cite{appendixA}.
The results, see top panel of Fig.~\ref{fig:entropy_fit}, are well fitted by the empirical functional form
\begin{equation}
    \overline{S_2(L;N)}=A(N) \log\left[ 1+\frac{B(N)}{\pi} \sin(\frac{\pi L}{N})\right] \,.
    \label{eq:entanglement-fit}
\end{equation}
Notice that, in the large $N$ limit, this functional form may alternatively represent a $\overline{S_2(L)}\sim L$ scaling, a $\overline{S_2(L)}\to 0$ scaling or a  $\overline{S_2(L)}\sim \log L$ scaling of the entropy, depending on the behaviour of the fit parameters $A(N)$ and $B(N)$.

%15
In the range of system sizes that we were able to explore~($N\leq200$) we observe that the parameter $B(N)$ converges to a finite constant as $N\to\infty$. Similarly, the product $ C(N)\equiv A(N)B(N)/N$ also converges to a constant $C$. This suggests the asymptotic behaviour $ \overline{S_2(L;N)} = C L + \mathcal{O}(1/N)$

in the thermodynamic limit.
In other words, we observe an entanglement scaling proportional to the volume $L$ of the considered subsystem, that is larger than the one both of a product state description and of a permutationally invariant model.

\begin{figure}
    \centering
    \includegraphics[width=0.5\textwidth]{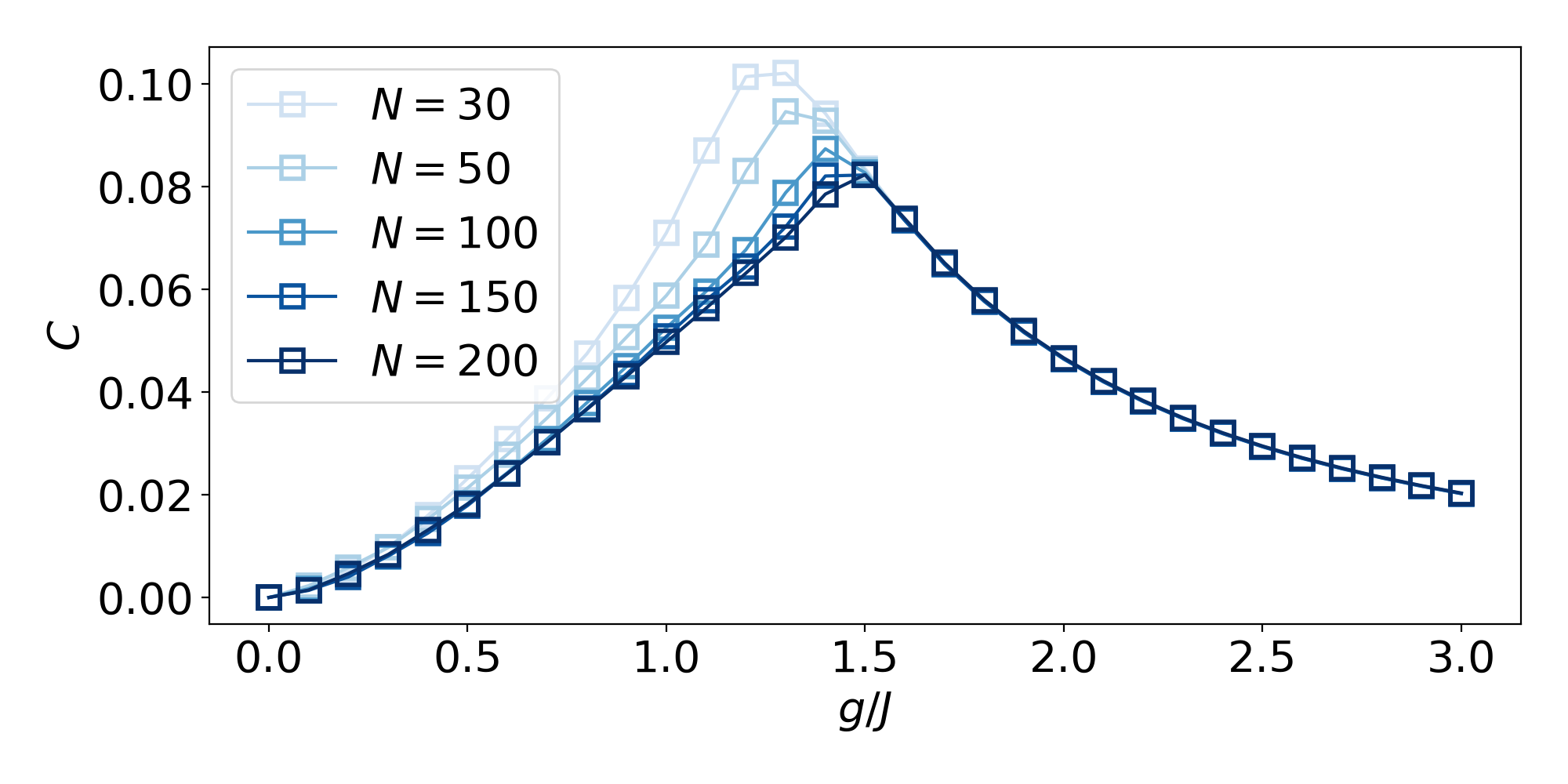}% there is also a big version with more Ns
    \caption{Coefficient $C(N)$ extrapolated from the R\'enyi entanglement entropy fit as a function of the transverse field $g$ at $h=0$ for different system sizes (different shades of blue). 
    }
    \label{fig:entropy}
\end{figure}

%16
Finally, we point out that the entanglement structure of the ground states appears to encode very clearly the phase transition of the model. More specifically, if we compute the fit coefficient $C(N)$ defined above %in~\eqref{eq:C},
as a function of the transverse field $g$ at $h=0$, we will see that this function develops, in the thermodynamic limit, a discontinuity in its derivative at the critical value $g_\mathrm{C}\approx 1.5 J$, as shown in Fig~\ref{fig:entropy}.

%17
\textit{Comparison to random weighted graph states} --
The form of the matrix $\boldsymbol{M}$, which appears to be distributed according to a GOE, suggests that the entanglement structure of the QSK ground states is encoded in a simple way in $\mathcal{V}\pqty{\boldsymbol{M}}$. To see this better, consider the set of states parametrized as
\begin{equation}
    \ket{\Psi\pqty{\boldsymbol{M}}} = \mathcal{V}\pqty{\boldsymbol{M}} \ket{+,\dots,+}\, ,
    \label{eq:wgs}
\end{equation}
where $\ket{+}=\frac{1}{\sqrt{2}}(\ket{\uparrow}+\ket{\downarrow})$. These are a subset of the full variational class~\eqref{eq:gcs} and, in the context of Quantum Information Theory, are referred to as \textit{weighted graph states}~\cite{hartmann_weighted_2007}. Let us then consider a random ensemble of such states constructed by drawing the matrix $\boldsymbol{M}$ from a GOE with variance $\overline{M_{nm}^2}=1/N$. 

%18
We can compute the average subsystem entanglement entropy $\overline{S_2(L)}$ for this ensemble of states, similarly to what we did for the ground states. We find that this entropy is fitted by the same functional form~\eqref{eq:entanglement-fit}, see bottom panel of Fig.~\ref{fig:entropy_fit}, and that the fit parameters $A(N)$ and $B(N)$ obey the same large $N$ scalings as in the ground state case. %~\eqref{eq:B-scaling} and~\eqref{eq:C}.
It is also possible to show analytically that the entanglement of these states must scale according to a volume law, as confirmed by these fits~\cite{appendixA}.

%19

%19
We conclude that the simple form~\eqref{eq:wgs}, where $\boldsymbol{M}$ is sampled from a GOE, exhibits the key entanglement features of the QSK ground states. It can be taken as a minimal example of this entanglement structure.

%20
Let us stress, however, that the actual ground states still contain more information than the states~\eqref{eq:wgs}. The state $\ket{\phi(\boldsymbol{x})}$ appearing in the variational ansatz~\eqref{eq:gcs} is in general not equal to $\ket{+,\dots,+}$. Rather, we observe that $\ket{\phi(\boldsymbol{x})}$ transitions from being $z$-polarized in the spin glass phase to being almost fully polarized in the $xy$-plane in the paramagnetic phase, encoding the information about the model's phase. Furthermore, the proportionality constant between $\overline{M_{nm}^2}$ and $1/N$ also shows a non-trivial dependence on $g$ and $h$.

\begin{figure}[t]
    \centering
    \includegraphics[width=0.5\textwidth]{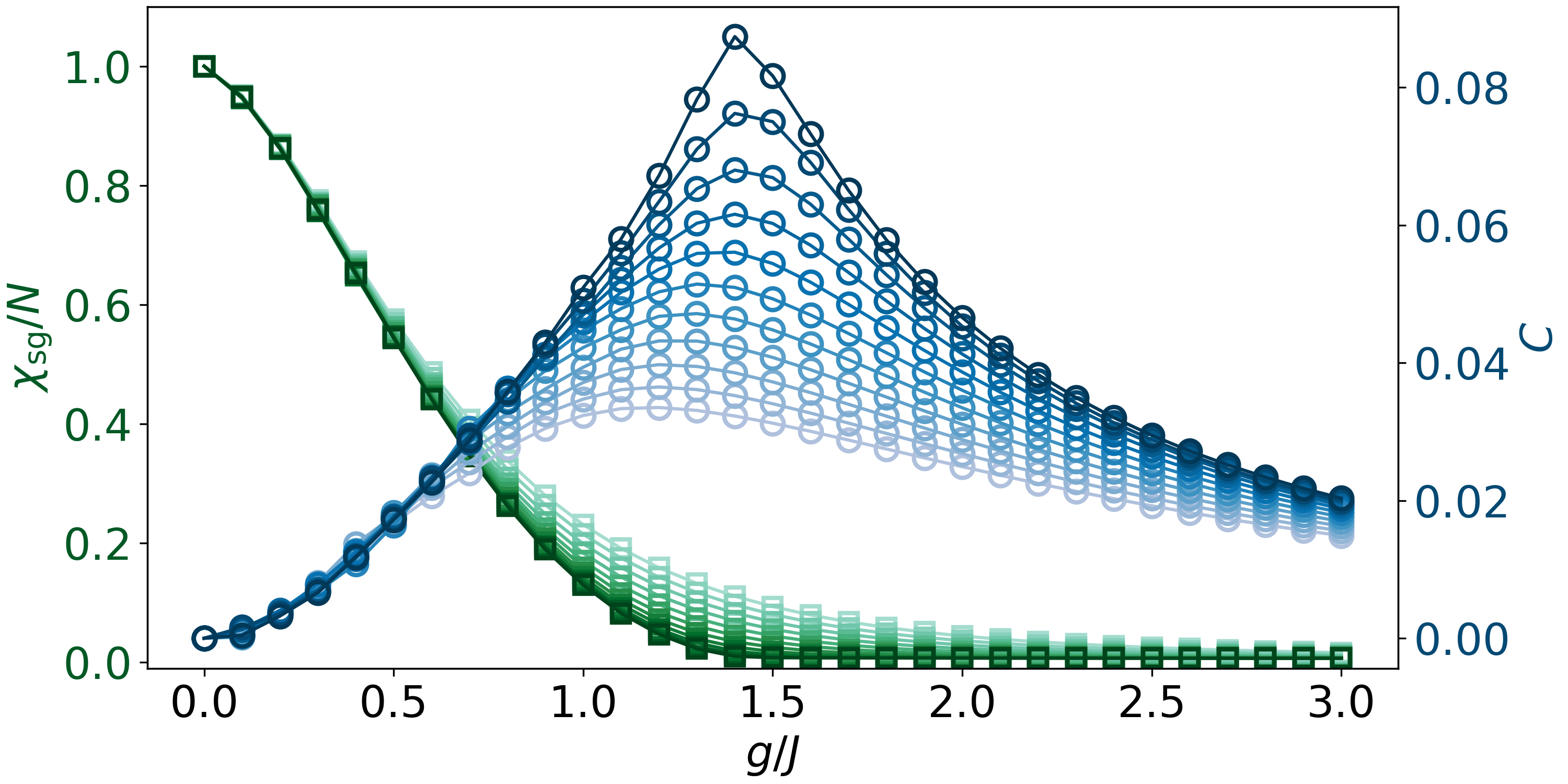}
    \caption{Spin glass susceptibility per site $\chi_{\mathrm{sg}}/N$~(green squares) and entropy coefficient $C$~(blue circles) as functions of $g$ for different values $h/J=0,\, 0.1,\, \dots 1$~(dark to light) at $N=150$. We observe that both functions develop a singularity typical of a phase transition only in the $h=0$ limit.}
    \label{fig:finite-h}
\end{figure}

%21
\textit{Phase transition at finite longitudinal fields} --
Another non-trivial feature of the QSK model which can be studied thanks to our method is the presence of a phase transition at $h>0$. It has been conjectured that the model's spin glass phase survives also for non-vanishing longitudinal fields $h$, suggesting the existence of a line of quantum phase transitions between the spin glass and paramagnetic phases that extends from the $g=g_C$, $h=0$ critical point into the $h>0$ plane (often referred to as the quantum de Almeida-Thouless line). This conjecture is however
based on not yet conclusive investigations of the stability of replica symmetry breaking (RSB) at zero temperature~\cite{young_stability_2017,Manai2021}.

%22
Our variational analysis can tackle this issue without making assumptions about RSB. Indeed, we can extend our analysis to variational ground states in the whole parameter space of the model, including $h>0$. We observe that all indicators of a phase transition vanish as soon as~$h>0$. 

%23
More specifically, the spin glass susceptibility $\chi_{\mathrm{sg}}$ becomes a smooth function of $g$ whenever $h>0$, no longer presenting the discontinuity in its derivative typical of a phase transition, even at large $N$. Similarly, the coefficient $C$ characterizing the entropy behaviour of the ground states clearly shows a singular behaviour at $h=0$ but not for finite $h$. These results are illustrated in Fig.~\ref{fig:finite-h}. In conclusion, our analysis was not able to identify any sign of the conjectured phase transition in the $h>0$ region. The discrepancy with previous results suggests this regime should be investigated further, especially in relation to RSB.

%24
\textit{Conclusion} --
We have shown that generalized atomic coherent states capture relevant properties of the ground states of the QSK model. The subset approximating the ground states contains a non-trivial entanglement structure, displaying a volume law, akin to the one of weighted graph states with random phase gates.

It is remarkable that the GCS resemble the QAOA ansatz~\cite{farhi2014quantum,Cerezo2021} for the QSK model, with one Ising interaction layer sandwiched between two product operators. In our case, however, the parameters of these layers do not necessarily correspond to those of the Hamiltonian. Our results can thus inspire other quantum computing variational eigensolvers for QSK-like models, building on top of the states~\eqref{eq:gcs}, something that deserves a more detailed research.

\textit{Acknowledgements} --
T.G acknowledges valuable discussions with Lorenzo Piroli and Nicola Pancotti.
J.I.C. and T.G. are supported by the ERC Advanced Grant QUENOCOBA under the EU Horizon2020 program (Grant Agreement No. 742102) and the German Research Foundation (DFG) under Germany's Excellence Strategy through Project No. EXC-2111-390814868 (MCQST).
The project is part of the Munich Quantum Valley, which is supported by the Bavarian state government with funds from the Hightech Agenda Bayern Plus.
P.M.S. has been financially supported by the Studienstiftung des Deutschen Volkes (German Scholarship Foundation).
T.S. is supported by the NSFC (Grants No. 11974363).
E.D. is supported by the ARO grant number W911NF-20-1-0163, AFOSR-MURI Photonic Quantum Matter award FA95501610323 and
the National Science Foundation through grant NSF EAGER-QAC-QSA, award number 2222-206-2014111. 
%Computational work reported in this paper was performed on the MPI PKS cluster.

P.M.S. and T.G. contributed equally to this work.

\FloatBarrier

% Bibliography
%%%%%%%%%%%%%%%%%%%%%%%%%

\bibliography{bibliography}

\begin{thebibliography}{54}
\bibitem [{fac()}]{factor2}%
  \BibitemOpen
  \href@noop {} {\bibinfo {title} {\corr{Note that in part of the literature a
  different convention is chosen when defining the
  Hamiltonian~\eqref{eq:QSK_model}, omitting the factor $1/2$ in the Ising
  coupling term. This difference has to be taken into account when comparing
  the critical transverse fields obtained by different authors.}}}\BibitemShut
  {Stop}%
\bibitem [{bum()}]{bump}%
  \BibitemOpen
  \href@noop {} {\bibinfo {title} {The fact that the point of maximal error
  moves with $n$ to larger $g$ implies that for some $g$'s in the interval
  $[1.0j, 1.5j]$ the error increases with $n$ at fixed $g$. we expect this
  increase to be temporary and to saturate somewhere below the maximal error
  value, which we observe to decrease with $n$.}}\BibitemShut {Stop}%
\end{thebibliography}

%  Appendix
%%%%%%%%%%%%%%%%%%%%%%%%%
\clearpage
%\appendix

\onecolumngrid
\begin{center}\large \textbf{Supplementary Material} \end{center}
\vspace{2mm}
\twocolumngrid

\subsection*{Variational Method}

In this appendix we will present the algorithm used to find the variational ground states for the generalized atomic coherent states of the form
\begin{equation}
    \ket{\Psi} = \mathcal{U}\pqty{\boldsymbol{y}} \mathcal{V}\pqty{\boldsymbol{M}} \ket{\phi\pqty{\boldsymbol{x}}}\, ,
    \label{eq:gcs-app}
\end{equation}
defined in the main body of this work.

\textit{Optimization Procedure} --
The GCS variational ground state is described by the parameters $\zeta^\star =\pqty{\boldsymbol{y}^\star,\, \boldsymbol{x}^\star,\, \boldsymbol{M}^\star}$ which minimize the energy $E\pqty{\zeta}=\bra{\Psi\pqty{\zeta}}H\ket{\Psi\pqty{\zeta}}$, where $H$ is the Hamiltonian of the system.
For optimization of the parameters we used the natural gradient descent~(natural GD) algorithm~\cite{hackl2020geometry}.
Like for standard gradient descent~(GD), natural GD starts at some initial state and iteratively updates the parameters, with an update based on the local structure, until a minimum in the energy is reached.
However, while the parameters in GD are updated in the direction $\boldsymbol{\mathcal{X}}$ of the energy gradient $\boldsymbol{\mathcal{X}} = - \boldsymbol{\nabla} E$, 
the direction for natural gradient descent is defined by
\begin{equation}
    \boldsymbol{g} \boldsymbol{\mathcal{X}} = - \boldsymbol{\nabla} E \, \label{eq:natural_gd}
\end{equation}
and encodes additional information on the curvature in terms of the local metric $g_{\mu\nu} = 2 \mathrm{Re}\braket{V_\mu}{V_\nu}$, where the tangential vectors $\ket{V_\mu}$ are specified below.
Natural GD in general leads to enhanced convergence compared to GD, however, like GD can get stuck in local, non-optimal minima of the energy.
In order to avoid this, we employ an adiabatic updating procedure. 
Thereby, we start at $g=0$, where the system is exactly described by a product state and apply the natural GD algorithm to a large number of random initial CS~(usually $10.000$) and use the state with minimal energy as the variational ground state for both CS and GCS at $g=0$.
Then, iteratively for increasing transverse field values $g>0$, we use $\zeta^\star \pqty{g } + \eta$, that is the variational ground state parameters of the point $g$ with some small perturbation $\eta$, as the starting point of the natural GD algorithm to find the optimal parameters $\zeta^\star\pqty{g+\delta g}$ for the point $g+\delta g$.

\textit{Analytical Expressions of desired Quantities} --
In order to perform the optimization procedure described above, we will need to compute quantities of the form
\begin{equation}
    \bra{\Psi} H \ket{\Psi}, \ \bra{\Psi} H \ket{V_\mu} ,\ \braket{V_\mu}{V_\nu} \, , \label{eq:quantities}
\end{equation}
corresponding to the energy of the state, the derivative of the energy with respect to the variational parameters and the local structure of the variational manifold, respectively.
The $\ket{V_\mu}$ are the so called tangential vectors, describing the change in the state $\ket{\Psi}$ upon infinitesimal change in the variational parameters
\begin{equation*}
    \ket{V_\mu} = \mathbb{Q}_\Psi \frac{\partial}{\partial\zeta^\mu} \ket{\Psi} \, ,
\end{equation*}
where we take the derivative with respect to the $\mu$'th variational parameter.
The projection $\mathbb{Q}_\Psi \ket{\phi} = \ket{\phi} - \braket{\Psi}{\phi} \ket{\Psi}$ removes all directions which lead only to a change in phase or amplitude of the state $\ket{\Psi}$, i.e. which would not change the physical state.

There are three different kinds of tangential vectors
\begin{equation}
    \begin{aligned}
        \ket{\mathcal{X}_n^a}  &= \mathbb{Q}_\Psi \mathcal{U}\pqty{\boldsymbol{y}}                     \mathcal{V}\pqty{\boldsymbol{M}}                                          \mathcal{U}\pqty{\boldsymbol{x}}  \pqty{i \sigma_n^a} \ket{\uparrow}^{\otimes N} \, , \\
        \ket{\mathcal{M}_{pq}} &= \mathbb{Q}_\Psi \mathcal{U}\pqty{\boldsymbol{y}}                     \mathcal{V}\pqty{\boldsymbol{M}} \bqty{\frac{-i}{4} \sigma_p^z \sigma_q^z} \ket{\phi\pqty{\boldsymbol{x}}} \, , \\ 
        \ket{\mathcal{Y}_m^b}  &= \mathbb{Q}_\Psi \mathcal{U}\pqty{\boldsymbol{y}}   \pqty{i \sigma_m^b} \mathcal{V}\pqty{\boldsymbol{M}}                                         \ket{\phi\pqty{\boldsymbol{x}}} \, .
    \end{aligned}\label{eq:tangential_vectors}
\end{equation}
corresponding to the three kinds of variational parameters $x_n^a$, $M_{pq}$ and $y_n^a$, respectively.

To evaluate the quantities~\eqref{eq:quantities} let us first observe that the adjoint action of the rotation unitaries $\mathcal{U}\pqty{\boldsymbol{x}}$, defined below equation~\eqref{eq:cs} of the main body of the paper, on a product of Pauli operators simply results in independently rotating each Pauli operator according to 
\begin{equation}
    \mathcal{U}\pqty{\boldsymbol{x}}^\dagger \sigma_n^a \sigma_m^b \dots \mathcal{U}\pqty{\boldsymbol{x}}= \pqty{ \sum_{a^\prime} R^{aa^\prime}_n \sigma_n^{a^\prime}}\pqty{ \sum_{b^\prime} R^{bb^\prime}_m \sigma_m^{b^\prime}} \dots \, \label{eq:adjoint_rep}
\end{equation}
with the orthogonal matrices $\boldsymbol{R}_n\pqty{\boldsymbol{x}} = \boldsymbol{R}_n\pqty{\boldsymbol{x}_n}$ depending only on the parameters for the $n$'th spin $x_n^a$, where $a=x,\,y,\,z$. Hence, for any of the quantities in equation~\eqref{eq:quantities}, we can take care of the action of $\mathcal{U}\pqty{\boldsymbol{y}}$ simply by rotating the Pauli operators that appear in $H$.

Let us now consider product operators $O=\bigotimes_{n=1}^N O_n$, where $O_n$ acts only on the $n$'th spin, since any operator is a linear combination of such product operators. 
Another direct consequence of equation~\eqref{eq:adjoint_rep} is that for two CS $\ket{\phi}$ and $\ket{\chi}$ the quantity $\bra{\phi}O\ket{\chi}$, where $O$ is an arbitrary product operator, factorizes into a product of $N$ single-spin terms 
\begin{equation}
    \bra{\chi} O \ket{\psi} = \prod_n \bra{\chi_n} O_n \ket{\psi_n} \, \label{eq:cs_exptvals}
\end{equation}
and is thus efficiently calculable.

Finally, in order to compute expectation values for GCS, the key observation is 
\begin{align}
    \mathcal{V}\pqty{\boldsymbol{M}}^\dagger \sigma_n^\alpha \mathcal{V}\pqty{\boldsymbol{M}} &= \sigma_n^\alpha \exp(\frac{\alpha i}{4} \sum_m M_{nm} \sigma_m^z)  \label{eq:gcs_relation} \\
    &\equiv \mathcal{O}_{n}^\alpha\pqty{\boldsymbol{M}} \, , \nonumber
\end{align}
for $\alpha\in\Bqty{+,\,-,\,0}$ and $\sigma^{\alpha=0}\equiv \sigma^z$.
Note that in what follows Greek superscripts $\alpha,\, \beta,\, \dots$ will refer to ${+,\, -,\, 0}$.
The relation~\eqref{eq:gcs_relation} is a direct consequence of the commutation relations $\comm{\sigma_n^\alpha}{\sigma^z_m}= \alpha \delta_{nm} \sigma^\alpha_n$.

Notice that the operator $\mathcal{O}_{n}^\alpha\pqty{\boldsymbol{M}}$ is a product operator.
Moreover, for multiple Pauli operators we can insert identities $\mathbb{1}= \mathcal{V}\mathcal{V}^\dagger$, such that
\begin{align*}
    \mathcal{V}^\dagger \sigma_n^\alpha \sigma_m^\beta \dots \mathcal{V} &= \mathcal{V}^\dagger \sigma_n^\alpha \mathcal{V} \mathcal{V}^\dagger \sigma_m^\beta \mathcal{V} \mathcal{V}^\dagger \dots \mathcal{V} \\
        &= \mathcal{O}_{n}^\alpha \mathcal{O}_{m}^\beta \dots 
\end{align*}
is again a product operator.

Thus, using the special relation~\eqref{eq:gcs_relation}, as well as the explicit form of the GCS~\eqref{eq:gcs-app} and the tangential vectors~\eqref{eq:tangential_vectors}, one immediately finds that the quantities~\eqref{eq:quantities} are simply sums of expectation values of product operators with respect to the CS part $\ket{\phi\pqty{\boldsymbol{x}}}$ of the GCS and can thus be computed efficiently.

Let us point out that the procedure described above for the computation of expectation values $\expval{\sigma_n^a \sigma_m^b \dots }_\Psi$ with respect to a GCS $\ket{\Psi}$ scales polynomially in the system size, but at the expense of scaling exponentially in the number of Pauli operators. However, for the present application one will have to compute expectation values of products of at most 4 Pauli operators, so this scaling does not pose a problem.

\subsection*{R\'enyi-$2$ Entropy}

In this appendix we will present a method that can be used to efficiently estimate numerically the second R\'enyi entropy of entanglement for states of the form
\begin{equation*}
    \ket{\Psi} = \mathcal{U}\pqty{\boldsymbol{y}} \mathcal{V}\pqty{\boldsymbol{M}} \ket{\phi\pqty{\boldsymbol{x}}}\, ,
\end{equation*}
that is GCS as defined in the main body of the paper.

\textit{Numerical computation} --
Let us consider a system of $N$ spins and a partition of the spins into two sets $A$ and $A^c$ constituted of $L$ and $N-L$ spins, respectively.
We are interested in computing the R\'enyi-$2$ entropy $S_2=-\log_2\pqty{q_A}$ of the reduced state $\rho_A=\tr_{A^c}\ket{\Psi}\bra{\Psi}$, where $q_A$ is the purity $q_A = \tr(\rho_A^2)$. Notice that the local unitaries contained in $\mathcal{U}\pqty{\boldsymbol{y}}$ do not modify this quantity in any way, so in what follows we will assume them to be all equal to the identity.

We will show that the quantity $q_A$ can be rewritten in terms of a sampling problem of a set of $L$ classical spin-$1$ variables, taking values $-1$, $0$ and $+1$. That is, we have
\begin{equation}
    q_A=\sum_{j_1,\dots,j_L=-1,0,+1} P_{1}(j_1) \cdots P_{L}(j_L) F(j_1,\dots,j_L)\,,
    \label{eq:entropy-as-sampling}
\end{equation}
for a certain function $F$ and certain probability distributions $P_n$. Thus, one can estimate $q_A$ by sampling configurations of the classical spins $\{j_n\}$ according to the product probability distribution $P_1\cdots P_L$ and computing the expectation value~\eqref{eq:entropy-as-sampling} as the mean value of $F$. 
To achieve an error $\epsilon$ on $q_A$ it is sufficient to sample $\sim 1/\epsilon^2$ configurations, rather than compute all the exponentially many terms in the sum~\eqref{eq:entropy-as-sampling}. Notice, that the entropy $S_2$ is invariant if one exchanges the sets $A$ and $A^c$, so we can always choose $A$ to be the smallest of the two.

To rewrite $q_A$ let us consider an ancillary system, also made up of $N$ spins and prepared to be in a copy of the state $\ket{\Psi}$. We will denote quantities relative to this ancillary system with primes. We then have
\begin{equation}
    q_A=\bra{\Psi,\, \Psi} \mathcal{S}_{AA^\prime} \ket{\Psi,\, \Psi}\,,
    \label{eq:purity-as-swap}
\end{equation}
where $\mathcal{S}_{AA^\prime}$ is the swap operator acting between the spins in $A$ and the corresponding ancillas in $A^\prime$.

Note that the terms in $\mathcal{V}\pqty{\boldsymbol{M}}$ that only connect spins within $A$ or within $A^c$ do not contribute to~\eqref{eq:purity-as-swap}. We can therefore replace $\mathcal{V}\pqty{\boldsymbol{M}}$ with 
\begin{equation*}
    \tilde{\mathcal{V}}\pqty{\boldsymbol{M}}=\exp(-\frac{i}{4}\sum_{n\in A^c} \sigma_n^z \sum_{m\in A} M_{nm} \sigma_m^z)\,.
\end{equation*}
We will also assume that $\ket{\phi\pqty{\boldsymbol{x}}}=\bigotimes_{n=1}^N \ket{\phi_n}$ with $\ket{\phi_n}=c_n^0 \ket{\uparrow} + c_n^1 \ket{\downarrow}$.

We can then write 
\begin{equation*}
    q_A=\bra{\phi\pqty{\boldsymbol{x}},\phi\pqty{\boldsymbol{x}}} \tilde{\mathcal{V}}^\dag \tilde{\mathcal{V}}^{\prime\dag} \mathcal{S}_{AA^\prime} \tilde{\mathcal{V}} {\tilde{\mathcal{V}}^\prime} \mathcal{S}_{AA^\prime} \ket{\phi\pqty{\boldsymbol{x}},\phi\pqty{\boldsymbol{x}}}\,,
\end{equation*}
where we exploited the fact that $\mathcal{S}_{AA^\prime}\ket{\phi\pqty{\boldsymbol{x}},\phi\pqty{\boldsymbol{x}}}=\ket{\phi\pqty{\boldsymbol{x}},\phi\pqty{\boldsymbol{x}}}$ to add an extra swap operator. We then act with the swap operators on $\tilde{\mathcal{V}} {\tilde{\mathcal{V}}^\prime}$ (exchanging system and ancilla operators in $A$) to obtain
\begin{align*}
    Q&\equiv \tilde{\mathcal{V}}^\dag \tilde{\mathcal{V}}^{\prime\dag} \mathcal{S}_{AA^\prime} \tilde{\mathcal{V}} {\tilde{\mathcal{V}}^\prime} \mathcal{S}_{AA^\prime}\\
    &=\exp(\frac{i}{4}\sum_{n\in A^c} (\sigma_n^z-\sigma_n^{z\prime}) \sum_{m\in A} M_{nm} (\sigma_m^z-\sigma_m^{z \prime}))\\
    &=\prod_{n\in A^c} \exp(\frac{i}{4} (\sigma_n^z-\sigma_n^{z\prime}) \sum_{m\in A} M_{nm} (\sigma_m^z-\sigma_m^{z \prime}))\\
    &\equiv\prod_{n\in A^c} Q_n \,.
\end{align*}
Each operator $Q_n$ has support on the single spin $n\in A^c$ and on all of $A$. We can therefore first take the expectation value of each $Q_n$ on the term $\ket{\phi_n,\phi_n}$ corresponding to the spin $n$ (of the system and of the ancilla) within the product state $\ket{\phi\pqty{\boldsymbol{x}},\phi\pqty{\boldsymbol{x}}}$. We easily find
\begin{equation*}
    \bra{\phi_n,\phi_n}Q_n\ket{\phi_n,\phi_n}=1-4p_n\sin^2\left[\frac{1}{4}\sum_{m\in A} M_{nm} (\sigma_m^z-\sigma_m^{z \prime})\right]\,,
\end{equation*}
where we set $p_n=|c_n^0|^2 |c_n^1|^2$.

We then proceed to take the expectation value of $\prod_{n\in A^c}  \bra{\phi_n,\phi_n}Q_n\ket{\phi_n,\phi_n}$ on the remaining part of the state $\ket{\phi\pqty{\boldsymbol{x}},\phi\pqty{\boldsymbol{x}}}$ corresponding to the subsystem $A$ which leads to the expression~\eqref{eq:entropy-as-sampling}, once we define
\begin{align*}
    P_m(0)&=|c_m^0|^4 + |c_m^1|^4 \\
    P_m(+1)=P_m(-1)&= |c_m^0|^2 |c_m^1|^2\,,
\end{align*}
which gives rise to a well-defined probability distribution. The function $F$ turns out to be
\begin{equation*}
    F(j_1,\dots,j_L)=\prod_{n\in A^c} \left[1-4p_n\sin^2\left(\frac{1}{2}\sum_{m\in A} M_{nm} j_m \right)\right]\,,
\end{equation*}
which can be evaluated efficiently for any configuration of $j$s. Note how the classical spin-$1$ variables $j_m$ emerge as the possible eigenvalues of the operators $(\sigma_m^z-\sigma_m^{z \prime})/2$.

This method for the computation of the second R\'enyi entropy appears to work very well when applied to the QSK ground states. In the numerical studies we performed for this work, we always observed a rapid convergence of the entropy values. Estimating the entropy of entanglement of a given state across a given partition of the system to within an error of $10^{-3}$ required a few seconds on a laptop computer in the worst case (\textit{i.e.} N=200 and L=N/2). Although we did not perform an in-depth analysis of this aspect, the data collected suggests that this computational cost scales polynomially in $N$.

\textit{Volume law scaling of entanglement} --
The form~\eqref{eq:entropy-as-sampling} of the purity can also be used to prove that the ensemble of random weighted graph states discussed around equation~\eqref{eq:wgs} of the main text of the paper must have a volume law scaling of the entanglement. 

For this, let us use the fact that $S_2$ is invariant under exchange of $A$ and $A^c$ to rewrite~\eqref{eq:entropy-as-sampling} as
\begin{align}
    S_2(L)&=-\log_2\left[ \sum_{\{j\}} P(j_{L+1}) \cdots P(j_N) F(j_{L+1},\dots,j_N) \right] \nonumber \\
    &=-\log_2\left[ \sum_{\{j\}} P(j_{L+1}) \cdots P(j_N) \prod_{n=1}^L f(X_n) \right]\,,
    \label{eq:entropy-as-function-of-f}
\end{align}
where $f(x)=1-\sin^2x$ and $X_n$ are random variables defined by
\begin{equation}
    X_n=\frac{1}{2}\sum_{m=L+1}^N M_{nm} j_m\,.
\end{equation}
Notice, that in the case of weighted graph states where we fix $\ket{\phi\pqty{\boldsymbol{x}}}=\ket{+\cdots +}$ the probability distributions $P_n$ are all the same for each $j_n$ and are given by $P(0)=1/2$, $P(\pm 1)=1/4$. 

The variables $X_n$ are the sum of a large number of independently distributed random numbers. By the Central Limit Theorem we can therefore assume that, in the limit of large $N$ (and fixed $L$), the variables $X_n$ are distributed according to normal distributions with mean and variance given by
\begin{align}
    \left<X_n\right>&=0\\
    \left<X_n^2\right>&=\frac{1}{8}\sum_{m=L+1}^N M_{nm}^2 \,,
\end{align}
where by $\left<\phantom{:}\cdot\phantom{:}\right>$ we denote averaging over the variables $j$. Note that each entry $M_{nm}$ of the matrix $\boldsymbol{M}$ is an independent identically normally distributed variable. We can therefore assume that in the large $N$ limit the sum $\sum_m M_{nm}^2$ will approximate the variance of $M_{nm}$. More precisely
\begin{equation}
    \sum_{m=L+1}^N M_{nm}^2 \approx (N-L) \, \overline{M_{nm}^2} = (N-L) \,\frac{1}{N}\,.
\end{equation}
It follows that $\left<X_n^2\right> \to 1/8$ for $N\to \infty$.

We can also assume that the variables $X_n$ are independently distributed. Indeed, their correlator is given by 
\begin{equation}
    \left< X_n X_m\right> = \frac{1}{8}\sum_{l=L+1}^N M_{nl}  M_{ml}\,.
\end{equation}
For $n\neq m$ this correlator has vanishing average with respect to the disorder of $\boldsymbol{M}$. Its variance is a function of $\overline{\boldsymbol{M}^2}$ and can be seen to decay as $1/N$.

From all these considerations we can conclude that equation~\eqref{eq:entropy-as-function-of-f} will ultimately reduce to
\begin{align}
    S_2(L)&= -\log_2\left<\prod_{n=1}^L f(X_n)\right> \\
    &= -\log_2 \prod_{n=1}^L \left<f(X_n)\right> \label{eq:factorisation}\\
    &= -\sum_{n=1}^L \log_2\left<f(X_n)\right>\\
    &= -\sum_{n=1}^L \log_2 \frac{1+e^{-2\left<X_n^2\right>}}{2} \label{eq:gaussian-integral}\,.
\end{align}
The factorisation in~\eqref{eq:factorisation} is valid only up to corrections containing the correlator $\left< X_n X_m\right>$, however we have seen that this correlator will go to zero at least as $1/N$ in the limit $N\to \infty$. In step~\eqref{eq:gaussian-integral} we have simply computed the average of $f(x)$ over a normally distributed variable with zero mean and variance $\left<X_n^2\right>$.

In the preceding paragraphs we have seen how, in the large $N$ limit, the variances $\left<X_n^2\right>$ actually neither depends on $n$ nor on the specific realization of $\boldsymbol{M}$, but rather all tend to $1/8$. We can therefore arrive at the result 
\begin{equation}
    \overline{S_2(L)} = L \left( -\log_2 \frac{1+e^{-\frac{1}{4}}}{2} \right) = C\,L
\end{equation}
which shows that the volume law entanglement entropy scaling holds in the limit of large $N$ and fixed $L$. The value of the constant $C$ that we have derived analytically here coincides numerically with the one that can be extracted from the functional fits discussed in the main body of this paper. 

A similar behaviour can be expected also in the case of the QSK ground states. There, however, the product state $\ket{\phi\pqty{\boldsymbol{x}}}$ has some structure which will make the distributions $P_n(j)$ depend on $n$. 
It follows, that the variances $\left< X_n^2\right>$ will also depend on $n$ in a non-trivial way. 
However, they will still be of order $1$ in the large $N$ limit and therefore the expression~\eqref{eq:gaussian-integral} remains extensive in $L$.

\subsection*{Additional observables}

\begin{figure}
    \centering
    \includegraphics[width=0.4\textwidth]{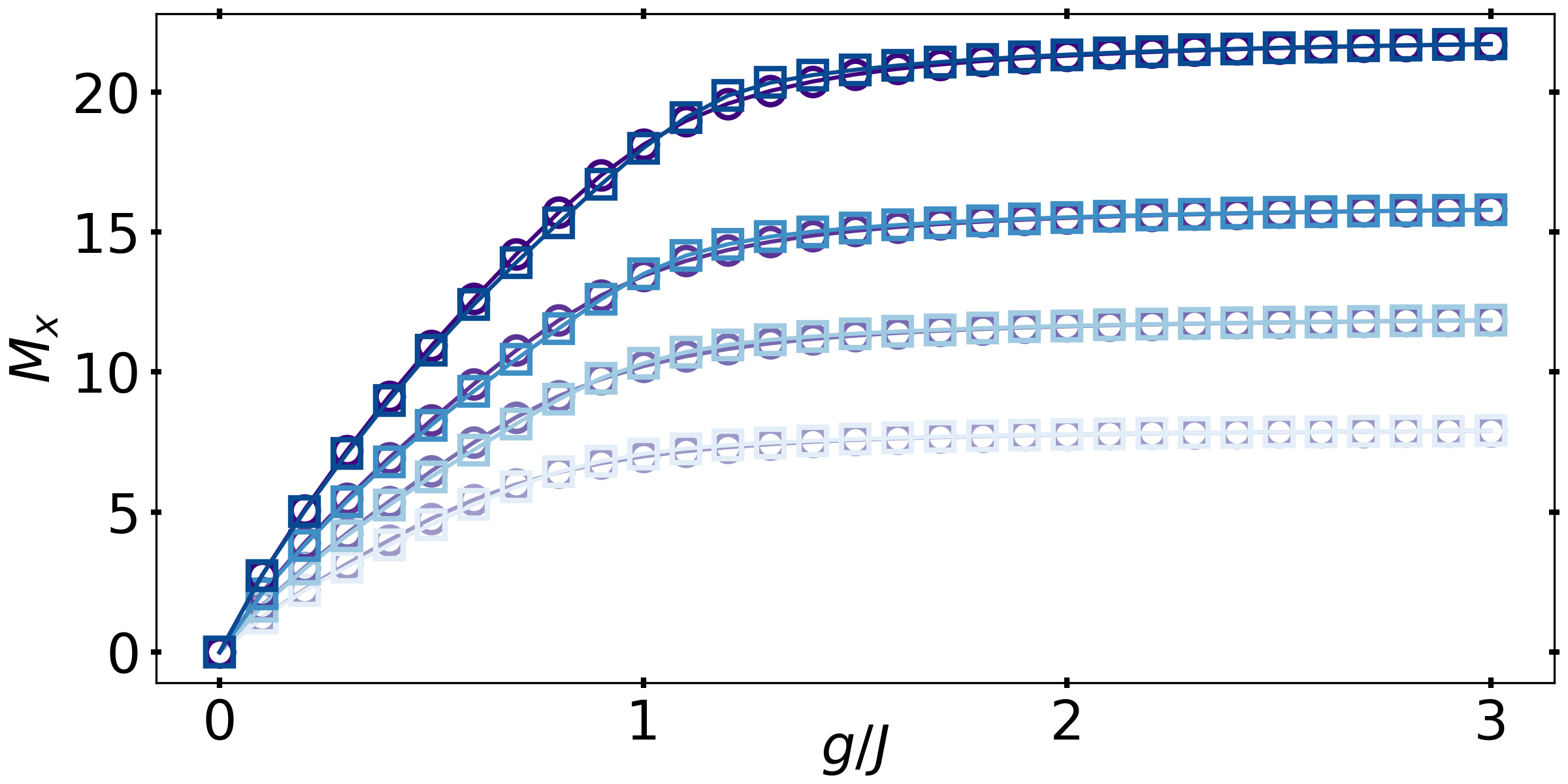}
    \caption{Total Magnetization $M_x$ as a function of the transverse field $g$ for ED~(purple circles) and GCS~(blue squares) for system sizes $N=8,\, 12,\, 16,\,22$~(from light to dark). 
    All data is for $h=0$ and averaged over $n_\mathrm{samples}=1000$ disorder realizations.}
    \label{fig:magnetization}
\end{figure}

In the main text we compare the variational groundstate result with the numerically exact groundstates based on the energy and spin glass susceptibility, see Fig.~\ref{fig:energy} and Fig.~\ref{fig:order_parameter}, respectively.
Here, we report on another quantity, the total disorder averaged magnetization $M_x= \sum_{n=1}^N \overline{\expval{\sigma_x}}$, see Fig.~\ref{fig:magnetization} of this Supplemental Material.

As for the other reported quantities we observe good quantitative agreement between the exact and variational results which does not deteriorate with increasing system size.

\end{document}